Electron Spin Resonance Scanning Tunneling Microscope of Non-Magnetic Molecules

Zion Hazan, Michael Averbukh and Yishay Manassen *: Department. of Physics, Ben Gurion University, P.O.Box 653, Beer Sheva 84105, Israel.
*Corresponding author: email address (manassen@bgu.ac.il).

**Abstract:** Electron Spin Resonance-Scanning Tunneling Microscopy (ESR-STM) of $C_{60}$ radical ion on graphene is a first demonstration on radical ions. ESR-STM signal at $g = 2.0 \pm 0.1$ was measured in accordance with macroscopic ESR of $C_{60}$ radical ion. The ESR-STM signal was bias voltage dependent, as it reflects the charge state of the molecule. The signal appears in the bias voltage which enables the ionization of the lowest unoccupied molecular orbital (LUMO) – creation of radical anion - and the highest occupied molecular orbital (HOMO) – creation of a radical cation - of the $C_{60}$ molecule when it deposited on graphene.

The linewidth provides information on the lifetime of the free radical. In several experiments, a $^{13}C$ hyperfine splitting was observed, and a change in the phase of the ESR-STM peak as a function of the speed of the magnetic field sweep was changed. This might be used to provide information on the relaxation times of the molecules. In parallel, ESR-STM signal at $g = 1.6 \pm 0.1$ was ascribed to Tungsten oxide ($W^{5+}$) at the tip apex, which is not bias voltage dependent.

The ability of ESR-STM to explore non paramagnetic molecules, significantly broadens the scope of this technique.

The attempt to detect and manipulate a single spin in a single molecule is a fundamental Challenge [1-5]. ESR-STM [6-12] is based on monitoring the tunneling current-current correlations. It is done in a static voltage V and a static magnetic field B, without an oscillating electromagnetic field. The experiments resulted in a signal at the expected Larmor frequency, which is sharp even at room temperature.

ESR-STM has been used to detect the hyperfine spectrum of a single spin [13,14] in SiC. Similarity to macroscopic ESR was demonstrated in the spectrum of silicon vacancy [14], showing hyperfine contributions from $^{29}$Si nuclei.

Recently, other related experiments were published. At low temperatures, a different type of ESR-STM was found using a spin polarized tip and RF irradiation [15-18]. Furthermore, single spin ENDOR (Electron nuclear double resonance) was performed [19] by applying a RF field at frequencies of the nuclear transitions and detecting changes in the intensity of the hyperfine line observed by ESR-STM; this facilitated measurements of the hyperfine coupling, the quadruple coupling and the nuclear Zeeman frequencies.

Several theoretical models for the ESR-STM phenomenon were published [20,21]. Both claim that the presence of two spins is necessary for observing ESR-STM. The model in [21] predicts that there must be at least two tunneling channels whose interference generates ESR-STM. There are experimental evidences that support this model.

Many atoms, defects or molecules are paramagnetic when they have an unpaired electron. There are many such free radicals but most molecules are not paramagnetic. Nevertheless, it is possible to put an extra electron in a nonparamagnetic species (to create a radical anion) or to take away an electron (to create a radical cation). Nonparamagnetic molecules can be ionized and these (sometimes unstable) radicals can be detected with ESR. In macroscopic ESR the ionized radicals are created by irradiation. Examples are irradiation of γ rays at $77^O$K (radiolysis) or photolysis (radiation with UV) [22]; Electron bombardment and trapping for example with noble gas matrix at cryogenic temperatures [23,24]. In the liquid phase, they are formed by oxidation reduction reaction. Many neutral molecules were shown to be converted into a paramagnetic ion in the appropriate conditions. The Benzene ion radical is a famous example [25]. Even biologically important molecule can form paramagnetic ions. Examples are many types of amino acids [26] and nucleotides [27]. The creation of radical anions can be also done with electrochemistry [28] by reduction. Applying cyclic polarimetry in which the voltage on the active electrode is ramped continuously in such a way that oxidation and reduction occurs consequently. The reduction reaction is performed inside the ESR cavity (when the ramped voltage is in the right range), and the concentration of the formed radical ion is high enough to be detected by ESR.

Radical ions can be produced with an STM by applying a bias voltage to remove or to put an electron in the species that is examined. STM is able to provide evidences of charging that results in the creation of paramagnetic spin centers: Examples are the F color centers formed by electron bombardment of MgO films. The defects (protrusions) are seen and the tunneling spectrum shows band gap states above the centers [29]. These defects were shown to be paramagnetic by macroscopic ESR. Sometimes, the magnetism can be revealed in the spectrum using the Kondo effect [30].

The creation of paramagnetic radical ions and the ability to observe their ESR spectra, will generalize the ESR-STM technique to nonparamagnetic species. Thus, it has the potential to become a comprehensive, nanometer scale, chemical analysis tool.

In this work we demonstrate ESR-STM on diamagnetic molecules. $C_{60}$ was the chosen molecule, and graphene as a substrate. The $C_{60}$ adsorbs in a well-defined configuration which result in a high-resolution image at room temperature. Graphene has structural, and electronic decoupling of absorbed molecules which ensures a better molecular ion's stability.

The experiments were carried out with a home-made STM operated in room temperature in ultrahigh vacuum (UHV) (base pressure $\leq 1.5 \times 10^{-10}\ Torr$). We used chemically etched tungsten tip ($W$) in all the experiments. To remove any hydrocarbons contamination from the surface, the graphene sample is annealed by gradually heating to ~250 ℃, while keeping the background pressure $\leq 5 \times 10^{-10}\ Torr$. Atomic resolution images (Fig. 1) were observed. A clear honeycomb monolayer structure was observed with interatomic distance of 1.4 Å, consistent with the expected values [31]. Commercially available $C_{60}$ powder (Sigma-Aldrich, 99.9% purity) was deposited on the Graphene substrate, from a Knudsen-cell type evaporator. The Knudsen-cell was gradually heated to $300\ °C$, the $C_{60}$ sublimation temperature. At a background pressure of $< 1.5 \times 10^{-9}\ Torr$, the $C_{60}$ evaporated for ~2 minutes to form sub-monolayer of molecules on the graphene. The $C_{60}$/graphene sample was annealed in ~150 ℃. Single $C_{60}$ molecules were observed with dimensions consistent with the literature value and previous STM experiments on $C_{60}$ (Fig. 1) [32].

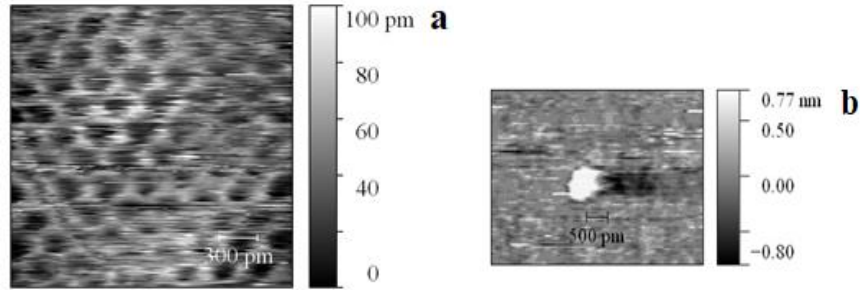

Figure 1: (a) Graphene atomic resolution image. (b) a single $C_{60}$ molecule.

Figure 2 displays the schematics of the ESR-STM setup. The STM tip was located above the identified molecule and the amplified $I_{AC}$ (The RF component) was recorded as a function of the magnetic field. The STM is equipped with solenoid that produce magnetic field sweep between $0 < B < 300\ G$. The magnetic field is applied parallel to the STM tip. The $I_{AC}$ was recorded by an HP spectrum analyzer operated at single frequency mode (zero sweep). The ESR-STM spectrum was observed as a function of the magnetic field and at different bias voltages..

We kept the measured frequency constant. That way, we eliminate any spurious noise, which might result from poor transmittance, due to mismatched impedances. The magnetic field was measured with Hall-probe chip placed near the STM tip.

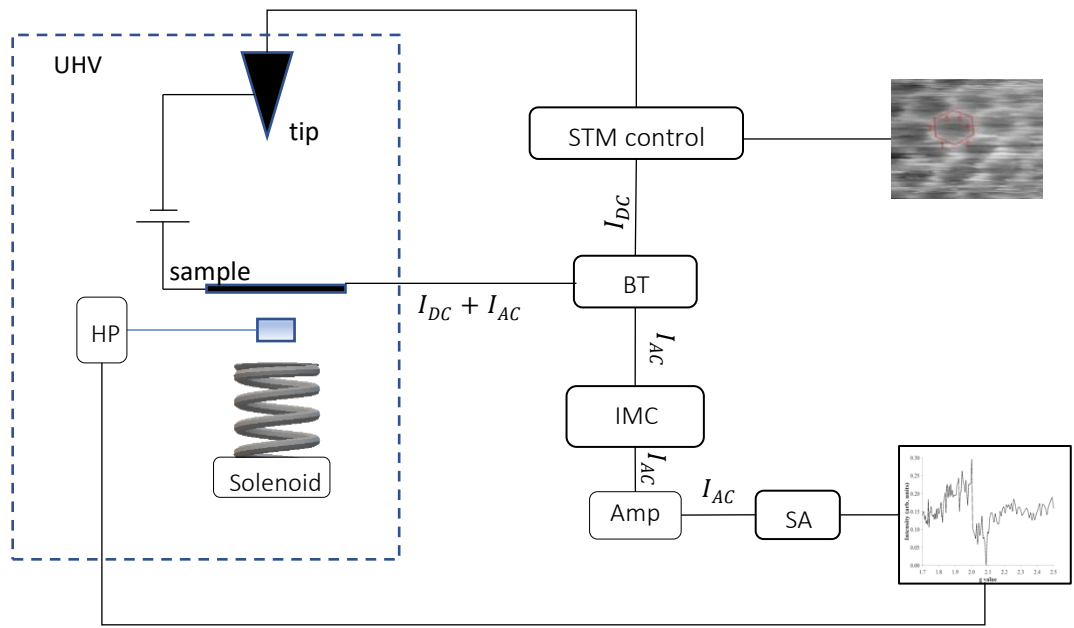

Fig. 2: Schematics of ESR-STM setup, including bias-tee (BT), impedance matching circuit (IMC), RF amplifier (Amp), *HP* spectrum analyzer (SA), solenoid for magnetic field sweep and Hall-probe (HP) to measure the magnetic field.

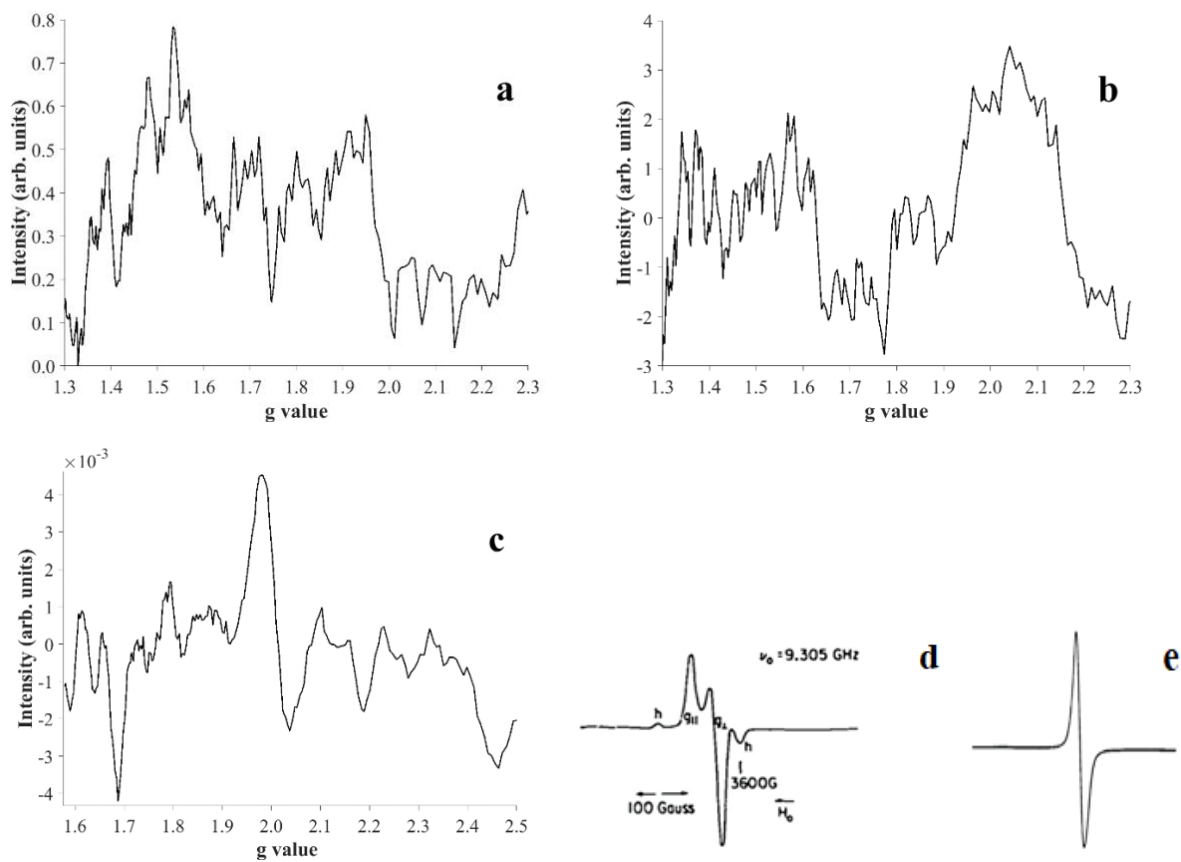

Fig. 3 (a) ESR – STM spectrum of $C_{60}$ on Graphene. Central frequency 300 MHz; Bias voltage (positive sample bias); 0.4 V; Tunneling current 0.4 nA. The vertical axis is with arbitrary units – also in (b) and (c). (b) Center frequency 300 MHz; Bias voltage 4.2 V; Tunneling current 0.4 nA. (c) Central frequency 400 MHZ; Bias voltage: -9V; Tunneling current: 0.5 nA. (d) and (e) are macroscopic ESR spectra of Tungsten and $C_{60,}$ respectively.

Our initial results are shown in Fig. 3 demonstrating the appearance of ESR-STM at different frequencies and different bias voltages. Fig 3 a-c shows that the results do not depend on the central frequency. It is noticeable, that the $C_{60}$ ESR signal disappears at lower bias voltages, suggesting ionization (both anions and cations) of the $C_{60}$ underneath the tip at higher bias voltage. The elevated intensity is observed at g values of: $g_1 \sim 1.6 \pm 0.1 \: and \: g_2 \sim 2.0 \pm 0.1$. Those values are compatible with macroscopic ESR of tungsten oxide ($W^{5+}$) [33-35] and $C_{60}$ molecular ions [36-40] respectively. The similarity to the macroscopic spectra is clear as shown in Fig. 3 d, e. (taken from [41, 36] respectively).

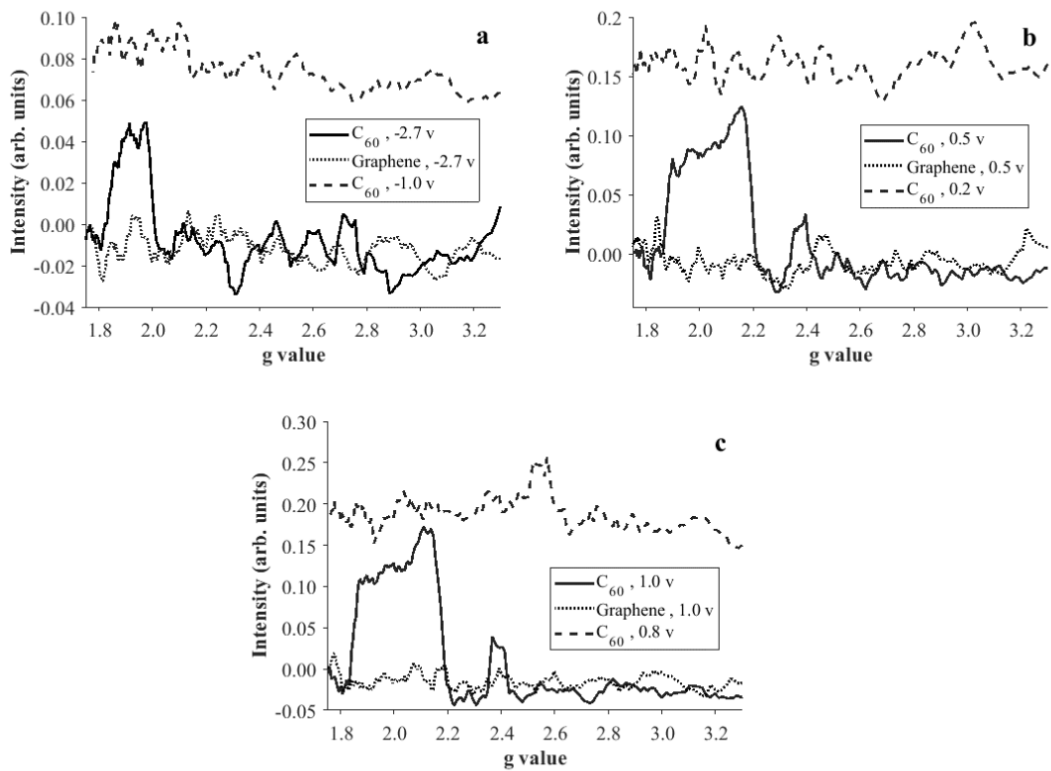

Fig. 4: ESR-STM of ion radicals at different bias voltages. In each spectrum it is compared with the signal of clean graphene and of the spectrum observed at lower bias voltage. Tunneling current in (a), (b) and (c) is 0.6 nA, while the central frequency is 385 MHz.

The appearance of ESR can be used to observe a lot of information on the measured molecule. The charging of the molecule (creating either a radical anion or cation) is found to happen when the tip - sample bias voltage is the same as the bias that is applied in tunneling spectroscopy when the HOMO (highest occupied molecular orbitals) and the LUMO (lowest unoccupied molecular orbital) appear in STS (scanning tunneling spectroscopy). As shown in [32], the HOMO orbital of $C_{60}$ on graphene is observed at a bias voltage of - 2.7 V and the LUMO orbital is at 0.8 V. In another publication [42] the LUMO orbital is shown to be also at 0.4 V, depends on the site on which the $C_{60}$ adsorbed on the graphene.

Figure 4 displays the measured ESR-STM spectra at a bias voltage when the spectrum appears. Each spectrum is compared with the null spectrum at the same bias voltage of a clean graphene surface and with the null spectrum observed at a slightly lower voltage on a $C_{60}$ molecule. All the spectra in Fig. 4 were done with central frequency of 380 MHz and tunneling current of 0.6 nA. The data (Fig. 4a) show that a broad ESR-STM signal appears at the correct energy of the HOMO orbital (due to the formation of a radical cation) at a bias voltage of -2.7 V, and in Fig. 4b, c it shows a signal close to the correct energy of the two LUMO orbitals - at 0.5 and 1 V. As was reported in [42] the two energies of the LUMO orbitals are due to different absorption sites.

It is quite easy to see that the relative (compared to the central frequency) linewidth of the ESR-STM peaks are larger than the macroscopic ESR data. As an example: the linewidth in Fig. 4b, c is 57 MHz, while in Fig. a is 30 MHz. It is reasonable, that the linewidth is an indication of the average time in which the free radical exists as a result of the flow of the tunneling electrons in and out of the molecule. This means that this time is $1.75 \cdot 10^{-8}$ and $3.3 \cdot 10^{-8}$ seconds respectively with-in the average time between two tunneling electrons is $3.6 \cdot 10^{-7}$ seconds, Of course, the relative linewidth can be improved by working with a much higher magnetic field.

The broad linewidth can hide a hyperfine coupling of $^{13}C$. However, in some other cases it can appear (Fig. 5a). It is recalled that the natural abundance of $^{13}C$ is 1.25% but the number of atoms in $C_{60}$ molecules is the reason why a hyperfine coupling can be observed. The splitting between the 2 neighbor peaks is about 12 G – quite close to the observed macroscopic hyperfine spectrum of the anion radical of $C_{60}$, which is 10 G [43]. There are several differences. The spectrum presented in Fig. 5a is a single molecule spectrum, and therefore the unequal intensity of the 2 peaks is not surprising and it is likely that the single molecule has a different hyperfine peaks intensity due to the fact that the measured molecule is under severe electric and stress fields which is not applied in a symmetric way by the STM tip.

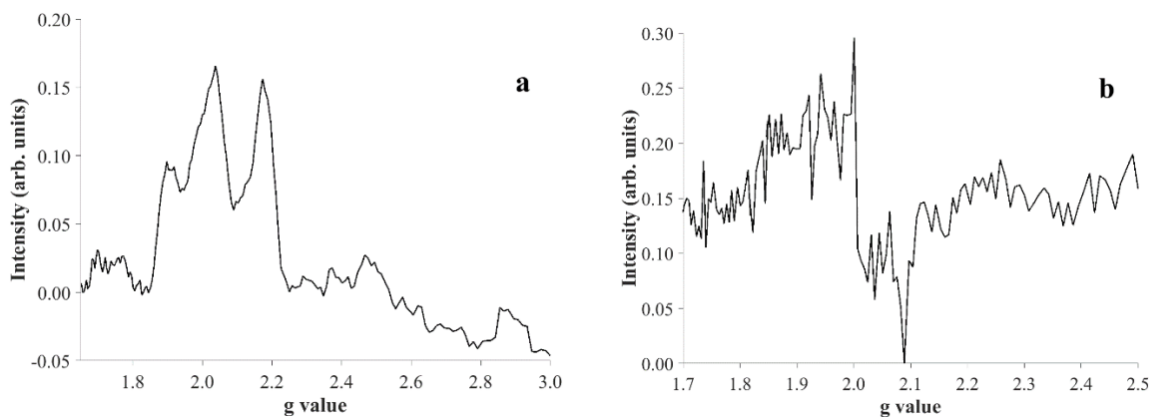

Fig. 5: (a) ESR-STM of $C_{60}$, hyperfine coupling of $^{13}C$ is observed. Bias voltage is 1 v, tunneling current is 0.6 nA and central frequency is 385 MHz. (b) A derivative spectrum at faster field sweep. Bias voltage is -2.7 v, tunneling current 0.6 nA and central frequency is 385 Mhz.

Finally, it is recalled [44-46] that when there is a rapid field sweep in macroscopic ESR, it results in a change of the phase of the signal as measured using magnetic field modulation. This was used to measure the spin lattice relaxation time ($T_1$). Surprisingly, a similar phenomenon was observed when an experiment similar to the one shown in Fig. 4a was found to change to a derivative signal (Fig. 5b) when the magnetic field sweep was faster by a factor of 2 (0.05 G/s and 0.1 G/s, respectively). It is obvious that the mechanism of the macroscopic spectrum is different. However, it might be that in this case also, this change is related to the spin lattice relaxation (which is in $C_{60}$ at room temperature it $5 \cdot 10^{-7}$ seconds [40]). This issue is currently in study.

The ionization of a molecule to create a radical ion and to measure its ESR-STM spectrum, has a significant potential due to the possibility to identify and to analyze unknown and non - magnetic molecules.

**Acknowledgements:**

This work was funded by the attract grant "NMR (1)". Additional funding was provided by the ISF – collaboration with China "ESR-STM study of individual radical ions at the single molecule level "and the DFG project "Magnetism of Vacancies and edge states in graphene probed by electron spin resonance and scanning tunnelling spectroscopy.

**References:**

[1] J. Wrachtrup, C. von Borczyskowski, J. Bernard, M. Orrit, and R. Brown, "Optical detection of magnetic resonance in a single molecule", *Nature*, **363**, 244 (1993). "Optically detected spin coherence of single molecule" ibid. *Phys. Rev. Lett.* **71**, 3565 (1993).

[2] J. Koehler, J. A. J. M. Disselhorst, M. C. J. M. Donckers, E. J. J. Groenen, J. Schmidt, and W. E. Moerner, "Magnetic resonance of a single molecular spin", *Nature* **363**, 242 (1993).

[3] D. Rugar, R. Budakian, H. J. Mamin, and B. W. Chui, "Single spin detection by magnetic resonance force microscope" *Nature* **430**, 329 (2004).

[4] J. M. Elzerman, P. Hanson, L. H. Willems van Beveren, B. Witkamp, L. M. K. Vandersypen, and L. P. Kouwenhoven, "Single shot read-out of an individual electron spin in a quantum dot" *Nature* **430**, 431 (2004).

[5] M. Xiao, I. Martin, E. Yablonovitch, and H. W. Jiang, "Electrical detection of the spin resonance of a single electron in a silicon field effect transistor", *Nature* **430**, 435 (2004).

[6] Y. Manassen, E. Ter-Ovanesyan, D. Shachal, and S. Richter, "Electron-Spin Resonance-Scanning Tunneling Microscopy Experiments on Thermally Oxidized Si(111)", *Phys. Rev. B* **48**, 4887 (1993).

[7] Y. Manassen, "Real-time response and phase-sensitive detection to demonstrate the validity of ESR-STM results" *J. Magn. Reson.* **126**, 133 (1997).

[8] Y. Manassen, I. Mukhopadhyay, and N. Ramesh Rao, "Electron-spin-resonance STM on iron atoms in silicon" *Phys. Rev. B* **61**, 16223 (2000).

[9] C. Durkan and M. E. Welland, "Electronic spin detection in molecules using scanning-tunneling microscopy assisted electron-spin resonance" *Appl. Phys. Lett.* **80**, 458 (2002).

[10] C. Durkan, "Detection of single electronic spins by scanning tunneling microscopy", *Contemp. Phys.* **45**, 1 (2004).

[11] T. Komeda and Y. Manassen, "Distribution of frequencies of a single precessing spin detected by scanning tunneling microscope", *Appl. Phys. Lett*. **92**, 212506 (2008).

[12] Y. Saaino, H. Isshiki, S.M.F. Shahed, T. Takaoka, and T. Komeda, "Atomically resolved Larmor frequency detection on Si(111)-7X7 oxide surface" *Appl. Phys. Lett.* **95**, 082504 (2009).

[13] A V. Balatsky, M. Nishijima and Y. Manassen, "Electron spin resonance scanning tunneling Microscopy" *Adv. Phys.* **61**,117 (2012).

[14] Y. Manassen, M. Averbukh and M. Morgenstern, "Analyzing multiple encounter as a possible

origin of electron spin resonance signals in scanning tunneling microscopy on Si(111) featuring C and O defects", *Surf. Sci.* **623**, 47 (2014).

[15] S. Mullegger, S. Tebi, A. K. Das, W. Schofberger, F. Faschinger, and R. Koch, "Radio Frequency Scanning Tunneling Spectroscopy for Single-Molecule Spin Resonance," *Phys. Rev. Lett.* **113**, 133001 (2014).

[16] S. Baumann, W. Paul, T. Choi, C. P. Lutz, A. Ardavan, and A. J. Heinrich, "Electron paramagnetic resonance of individual atoms on a surface", *Science* **350**, 417 (2015).

[17] P. Willke et al., "Probing quantum coherence in single-atom electron spin resonance", Sci. Adv. 4, eaaq1543 (2018).

[18] T. S. Seifert, S. Kovarik, C. Nistor, L. Persichetti, S. Stepanow, P. Gambardella, "Single-atom electron paramagnetic resonance in a scanning tunneling microscope driven by a radio-frequency antenna at 4°K" *Phys. Rev. Research* **2**, 013032 (2020).

[19] Y. Manassen, M. Averbukh, M. Jbara, B. Siebenhofer, A. Shnirman, B. Horovitz, "Fingerprints of single nuclear spin energy levels using STM ENDOR", *J. Magn. Reson,* **289**, 107 (2018).

[20] A. V. Balatsky, Y. Manassen, and R. Salem, "Exchange-based noise spectroscopy of a single precessing spin with scanning tunnelling microscopy" *Phil. Mag. B* **82**, 1291 (2002); *Phys. Rev. B* **66**, 195416 (2002).

[21] B. Horovitz and A. Golub, "Double Quantum Dot scenario for spin resonance in current noise", *Phys. Rev. B* **99**, 241407(R) (2019).

[22] T. Gillbro, " ESR and structure of sulfur centered radicals and radical ions. γ irradiated Dimethyl di sulfide and methane thiol single crystals at 77°K" *Chem. Phys.* **4** 476 (1974).

[23] T. Shida, E. Haselbach and T. Bally, "Organic radical ions in rigid systems" *Acc. Chem. Res.* **17**, 180 (1984).

[24] L. B Knight Jr. "ESR investigations of molecular cation radicals in neon matrices at 4OK: Generation, trapping and ion neutral reactions" *Acc. Chem. Res.* **19**, 313 (1986).

[25] J. R. Bolton, " $^{13}$C hyperfine splitting in the benzene negative ion" *Mol. Phys* **6**, 219 (1963).

[26] M. Ogawa, K. Ishigure and K. Oshima, "ESR study of irradiated single crystals of amino acids -1" *Radiat. Phys. Chem.* **16**, 281 (1980).

[27] A. Mueller, "Quantitative E.S.R.-measurements of radiation-induced radicals in nucleotides" *Internat. J Radiat. Biol.* **8**, 131 (1964).

[28] A. H. Maki and D. H. Geske, "Electron-Spin Resonance of Electrochemically Generated Free Radicals. Isomeric Dinitrobenzene Mono Negative Ions", *J. Chem. Phys.* **33**, 825 (1960).

[29] M. Sterrer, E. Fischbach, M. Heyde, N. Nilius, H.-P. Rust, Th. Risse, and H.-J. Freund, "Electron Paramagnetic Resonance and Scanning Tunneling Microscopy Investigations on the Formation of $F^+$ and $F^0$ Color Centers on the Surface of Thin MgO(001) Films" *J. Phys. Chem.* B, **110**, 8665 (2006).

[30] T. Komeda, H. Isshiki, J. Liu, Y. –F Zhang, N. Lorente, K. Katoh, B. K. Breedlove and M. Yamashita, "Observation and electric current control of a local spin in a single – molecule magnet", *Nature Comm.* **2** 217 (2011).

[31] H. Yang, A.J. Mayne, M. Boucherit, G. Comtet, G. Dujartin, and Y. Kuk. "Quantum interference channeling at graphene edges." *Nano letters* **10** 943 (2010).

[32] J. Cho, J. Smerdon, L. Gao, Ö. Süzer, J. R. Guest, and N. P. Guisinger. "Structural and Electronic decoupling of $C_{60}$ from epitaxial graphene on SiC." *Nano letters* **12** 3018 (2012).

[33] F. N. Studer, N. Rih and B. Raveau "Mixed Valence Tungsten Phosphate Glasses: ESR and Diffuse Reflectance Investigation" *J. Non. Crystal. Sol.* **107**, 101 (1988).

[34] R. R. Rakhimov *et. Al*. "Electron Paramagnetic Resonance Study of $W^{5+}$ pairs in Lithium – Tungsten Phosphate glasses" *J. Phys. Chem. B* **104**, 10973 (2000).

[35] A. Punnoose and M. S. Seehra, "ESR Observation of $W^{5+}$ and $Zr^{3+}$ States in $Pt/WO_x/ZrO_2$ Catalysts *Catalysis Lett.* **78** 157 (2002).

[36] P. Parimal *et al. "*Artifacts in the Electron Paramagnetic Resonance Spectra of $C_{60}$ Fullerene Ions: Inevitable $C_{120}O$ Impurity" *J. Am. Chem. Soc*. **124**, 4394 (2002).

[37] P. N. Keizer et al. "EPR Spectrum of Carbon Ion ($C_{60}$) Trapped in Molecular Sieve 13X" *J. Phys. Chem.* **95**, 7117 (1991).

[38] A. Colligiani and C. Taliani, "ESR Study on the Doublet and Triplet Species in Pristine $C_{60}$ Fullerene Powder" *Chem. Mater.* **6**, 1633 (1994).

[39] S. K. Hoffmann, *et al*. "Electron spin echo and EPR studies of paramagnetic centers in polycrystalline $C_{60}$." *Sol. Sta. Commun.* **93**, 197 (1995).

[40] G. G. Fedoruk, "Paramagnetic relaxation kinetics of the cation radical $C_{60}^+$ in $C_{60}$ powder." *Phys. Sol. Stat.* **42** 1182 (2000).

[41] M. V Krishnamurthy, "The determination of tungsten by electron paramagnetic resonance spectrometry." *Anal. Chim. Acta.* **124** 211 (1981).

[42] J. Lu, P. Shar, E. Yeo, Y. Zheng, Z. Yang, Q. Bao, Ch. K. Gan, and K. P. Loh, "Using the Graphene *Moiré* Pattern for the Trapping of $C_{60}$ and Homoepitaxy of Graphene" *ACS Nano* **6** 944, (2012).

[43] P. N. Keizer, J. R. Morton, K. F. Preston and A. K. Sugden, "EPR Spectrum of $C_{60}^-$ Trapped in Molecular Sieve 13X" *J. Phys. Chem.* **95**, 7117 (1991).

[44] F. Bloch, "Nuclear Induction" *Phys. Rev.* **70**, 460 (1946).

[45] A. M. Portis, "Rapid Passage Effects in Electron Spin Resonance" *Phys. Rev.* **100**, 1219 (1955).

[46] J. S. Hyde, "Magnetic Resonance and Rapid Passage in Irradiated LiF" *Phys. Rev.* **119**, 1483, (1960).